\documentclass[aps,amssymb,showpacs,twocolumn]{revtex4}
\usepackage{graphicx}
\usepackage{slashed}

\begin{document}
\title{ Dimuonium $(\mu^+\mu^-)$ Production in a Quark-Gluon Plasma }
\author{ Yibiao Chen and Pengfei Zhuang }
\affiliation{ Department of Physics, Tsinghua University, Beijing
100084, China}
\date{\today}

\begin{abstract}
We study dimuonium $(\mu^+\mu^-)$ production in the quark-gluon
plasma created in relativistic heavy ion collisions. The production
is controlled by the process $q\bar q\rightarrow
(\mu^+\mu^-)g$, and the dimuonium motion in the plasma is
described by a transport equation. While the electrodynamics
dominated dimuonium yield is not high enough, the transverse energy
distribution carries the information of the plasma at RHIC and LHC
energies.
\end{abstract}

\maketitle

The observation of positronium $(e^+e^-)$~\cite{Deutsch} and the
clarification of the leptonic nature of muons lead to a natural
question that whether there are ``true muonium''~\cite{Hughes1} or
``dimuonium''~\cite{Malenfant, Karshenboim} states $(\mu^+\mu^-)$.
Note that the terminology ``muonium" has been signed to the bound
state $(\mu^+ e^-)$, which has been discovered in
1960~\cite{Hughes2}. Since dimuonium is an ideal system of quantum
electrodynamics (QED), many of its properties are theoretically
predicted~\cite{Hughes1,Bilen'kii} with standard calculations of
QED. For instance, dimuonium is one of the most compact QED systems
with Bohr radius $512$ fm, the binding energy of the ground state
($1s$) is $1.4$ KeV, and the life times for the states $^1S_0$ and
$^3S_1$ are respectively $0.602$ ps (decay to $\gamma\gamma$) and
$1.81$ ps (decay to $e^+e^-$), which are both much shorter than the
life time of muon ($2.197$ $\mu$s) and thus the muon weak decay can
be ignored in the production of dimuonium. Many dimuonium production
mechanisms have been proposed, like direct $\mu^+\mu^-$
collision~\cite{Hughes1}, $e^+e^-\rightarrow
(\mu^+\mu^-)$~\cite{Moffat}, $\pi^-
p\rightarrow(\mu^+\mu^-)n$~\cite{Bilen'kii}, $\gamma
A\rightarrow(\mu^+\mu^-)A$~\cite{Bilen'kii} and $eA\rightarrow
e(\mu^+\mu^-)A$~\cite{Holvik}, where $A$ stands for a nucleus.
Dimuoniums can also be produced in heavy ion collisions through a
pure electromagnetic process $A_1A_2\rightarrow A_1A_2(\mu^+\mu^-)$,
where the mechanism is the fusion of coherent photons emitted from
the nuclei~\cite{Ginzburg}. Recently, the possibility of dimuonium
production at modern electron-positron colliders is
investigated~\cite{Brodsky}. However, dimuonium has not yet been
experimentally discovered.

It is generally believed that there is a quantum chromodynamics
(QCD) phase transition in hot and dense nuclear matter, which is
related to the deconfinement process in moving from a hadron gas to
a quark-gluon plasma (QGP)~\cite{Karsch}. The realization of such a
phase transition in laboratories can only be through relativistic
heavy ion collisions~\cite{Qm2011}. Considering the decreasing
temperature and density during the rapid expansion of the fireball
formed in the collisions, the QGP can not be measured directly in
the cold and dilute final state, and one needs signatures to
identify the QGP formation. Since leptons interact with particles
only through electromagnetic channel, the leptons produced in the
QGP carry the information of the plasma and will not lose it when
they pass through the collision region to reach the detectors.
Taking into account the fact that the lepton number in the fireball
increases with the temperature of the plasma, the probability to
produce a dimuonium state in high energy nuclear collisions should
be much larger than that in elementary electron-positron and
nucleon-nucleon collisions. In this paper we study the dimuonium
production in a hot QGP and calculate the production rate in high
energy nuclear collisions at relativistic heavy ion collider (RHIC)
and large hadron collider (LHC)~\cite{Qm2011}. Since the dimuonium
life time is much longer than the QGP life time (about several
fm/c), the dimuonium decay and dissociation in the QGP can safely be
neglected. In our calculation we will also ignore the dimuonium
production in the initial state and in the hadron gas, since their
contribution should be much smaller in comparison with the
production in the QGP.

There are two kinds of production processes in QGP for the dimuonium
states $(\mu^+\mu^-)$, one is from the lepton pair annihilation and
the other is from the quark-antiquark pair annihilation. While the
QCD processes in vacuum at RHIC and LHC energies may still be
non-perturbative, the medium effects will largely reduce the
effective coupling constant $\alpha_s$ ($\sim 0.3$)~\cite{Kaczmarek}
and may make the perturbative analysis available. By counting the
number of vertexes and considering the phase space suppression, one
can simply estimate the order of the dimuonium production cross
sections in different processes. The cross sections of the two
processes at leading order, $l^+l^-\rightarrow(\mu^+\mu^-)$ and
$q\bar{q}\rightarrow(\mu^+\mu^-)$, are of the order of $\alpha^2$,
with $\alpha$ being the electromagnetic coupling constant. Due to
the requirement of energy conservation, only the initial states in a
narrow energy window around the binding energy $1.4$ KeV contribute
to the production processes, and therefore the both cross sections
at leading order become very small. The next leading process is
$q\bar{q}\rightarrow(\mu^+\mu^-)g$ with the cross section of the
order of $\alpha_s\alpha^2$. The subsequent processes are
$l^+l^-(q\bar{q})\rightarrow(\mu^+\mu^-)\gamma$ and
$q\bar{q}\rightarrow(\mu^+\mu^-)gg$. The former is of the order of
$\alpha^3$ and the latter is of the order of $\alpha_s^2\alpha^2$.
Since the latter includes three particles in the final state, the
process is strongly suppressed in comparison with the former. Other
processes are higher order contributions. Therefore, we consider in
the following the main process $q\bar{q}\rightarrow(\mu^+\mu^-)g$
for the dimuonium production in the QGP. The contribution from the
other processes is at least ten times smaller from the simple vertex
counting.

\begin{figure}[!hbt]
\centering
\includegraphics[width=0.45\textwidth]{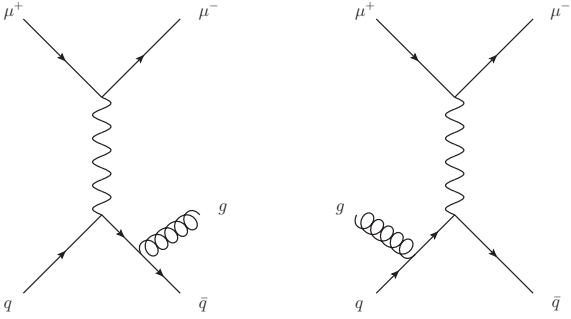}
\caption{The Feynman diagrams at tree level for the main dimuonium
production process $q\bar q\rightarrow (\mu^+\mu^-)g$ in QGP.}
\label{fig1}
\end{figure}

The Feynman diagrams at tree level for the main production process
$q\bar{q}\rightarrow(\mu^+\mu^-)g$ are shown in Fig.\ref{fig1}.
Since strange quarks are much heavier than light quarks, the
amount of strange quarks in QGP is less than $10\%$ of the amount of
light quarks at RHIC and LHC energies, and we consider only light quarks
in the calculation ($N_f=2$), and their current mass is set to be
zero. The scattering amplitude $\overline {\cal M}(q_1q_2\rightarrow
p_1p_2k)$ for the process $q\bar q\rightarrow \mu^+\mu^- g$ can be
written as
\begin{eqnarray}
\label{amplitude1} \overline{\cal M}&=&gQe^2\overline
v(q_2)\Big[t\gamma^\mu{1\over \slashed {q}_2-\slashed {k}}\gamma^\nu
+\gamma^\nu{1\over \slashed {q}_1-\slashed
{k}}t\gamma^\mu\Big]\nonumber\\
&\times&u(q_1){g_{\nu\rho}\over (p_1+p_2)^2}\overline
u(p_1)\gamma^\rho v(p_2)\epsilon_\mu^*(k),
\end{eqnarray}
where $q_1, q_2, p_1, p_2$ and $k$ are respectively the momenta of
$q$, $\bar q$, $\mu^+$, $\mu^-$ and $g$, $Q$ is the quark charge
number, $t$ is the color matrix, $g$ is related to the effective
coupling constant via the definition $\alpha_s=g^2/(4\pi)$, $u, v,
\overline u$ and $\overline v$ are Dirac spinors, and $\epsilon_\mu$
is the gluon polarization vector satisfying
$\epsilon_\mu^*(k)\epsilon_\nu(k)=-g_{\mu\nu}+k_\mu k_\nu/m_g^2$
with thermal gluon mass $m_g=2gT/3$ in the deconfinement
phase~\cite{Pisarski}. For simplicity, we have dropped in
(\ref{amplitude1}) the flavor, color and spin indices of quarks and
gluons.

Introducing the total and relative momenta $p=p_1+p_2$ and
$p'=(p_1-p_2)/2$ of $\mu^+$ and $\mu^-$, the amplitude
$\overline{\cal M}(q_1q_2\rightarrow p_1p_2k)$ can be expressed as
$\overline{\cal M}(q_1q_2\rightarrow pp'k)$. Since the bound state
$(\mu^+\mu^-)$ is non-relativistic, we can compute its production
amplitude ${\cal M}(q_1q_2\rightarrow pk)$ by integrating out the
relative momentum $p'$ in the center-of-mass frame of the
dimuonium~\cite{Peskin},
\begin{eqnarray}
\label{amplitude2}
{\cal M}&=&\sqrt{\frac{2}{m_d}}\int \frac{d^3
{\bf p}'}{(2\pi)^3}\overline{\cal M}(q_1q_2\rightarrow
pp'k)\psi^*({\bf p}'),\nonumber\\
\psi({\bf p}')&=&\frac{8\sqrt{\pi}a^{3/2}_0}{(1+a^2_0{\bf p}'^2)^2},
\end{eqnarray}
where $m_d$ is the dimuonium mass, and $\psi({\bf p}')$ is the
relative wave function for the ground state in momentum space with
$a_0$ being the Bohr radius. From the known scattering amplitude, we
can calculate the dimuonium production cross section $\sigma_{q\bar
q}^{(\mu^+\mu^-)g}(s)$ as a function of $s=(q_1+q_2)^2$ and the
transition probability
\begin{equation}
\label{transition}
W^{(\mu^+\mu^-)g}_{q\bar q}(s)=\frac{16\pi
s^2}{\sqrt{(s-m_d^2-m_g^2)^2-4m_d^2m_g^2}}\sigma^{(\mu^+\mu^-)g}_{q\bar
q}(s).
\end{equation}

The medium created in high-energy nuclear collisions evolves
dynamically. In order to extract information about the medium by
analyzing the dimuonium distributions, both the hot and dense medium
and the dimuonium production process must be treated dynamically. In
this paper, we treat continuous dimuonium production in QGP
self-consistently, including hydrodynamic evolution of the QGP.
Since dimuonium is a pure electromagnetic system, it can not be
thermalized with the medium which is governed by strong
interactions. Thus its phase space distribution should be controlled
by a transport equation. The transport equation should then be
solved together with the hydrodynamic equation which characterizes
the space-time evolution of the QGP.  Considering that dimuonium is
a heavy bound state, we use a classical Boltzmann-type transport
equation to describe its evolution. The distribution $f({\bf
p}_t,y,{\bf x}_t,\eta,\tau|{\bf b})$ as a function of transverse
momentum ${\bf p}_t$ and coordinate ${\bf x}_t$, longitudinal
rapidity $y$ and space-time rapidity $\eta$ and proper time $\tau$
at fixed impact parameter ${\bf b}$ in a heavy ion collision is
characterized by the equation
\begin{equation}
\label{transport}
\left[\cosh(y-\eta)\frac{\partial}{\partial\tau}+\frac{1}{\tau}\sinh(y-\eta)\frac{\partial}{\partial\eta}+{\bf
v}_t\cdot\nabla_t\right]f=\beta
\end{equation}
with the dimuonium transverse velocity ${\bf v}_t={\bf p}_t/E_t$ and
transverse energy $E_t=\sqrt{m_d^2+{\bf p}_t^2}$. Because the life
time of dimuonium is much longer than the life time of QGP, the
decay of dimuonium is ignored. Moreover, since dimuonium does not
participate in strong interactions, the generated dimuonium
interacts with QGP only electromagnetically, the dissociation can
then be neglected too. Therefore, we consider only the gain term
$\beta({\bf p}_t,y,{\bf x}_t,\eta,\tau|{\bf b})$ on the right hand
side of the transport equation,
\begin{eqnarray}
\label{rate}
\beta&=&\frac{1}{2E_t}\int\frac{d^3{\bf
k}}{(2\pi)^32E_g}\frac{d^3{\bf
q}_1}{(2\pi)^32E_q}\frac{d^3{\bf q}_2}{(2\pi)^32E_{\bar q}}W^{(\mu^+\mu^-)g}_{q\bar q}(s)\nonumber\\
&\times&f_qf_{\bar q}(1+f_g)(2\pi)^4\delta^{(4)}(p+k-q_1-q_2),
\end{eqnarray}
where $E_q=\sqrt{m_q^2+{\bf q}_1^2}$, $E_{\bar q}=\sqrt{m_q^2+{\bf
q}_2^2}$ and $E_g=\sqrt{m_g^2+{\bf k}^2}$ are respectively quark,
anti-quark and gluon energies, and $f_q$, $f_{\bar q}$ and $f_g$ are
the thermal distributions for quarks and gluons,
$f_q=1/\left(e^{q_1^\mu u_\mu/T}+1\right)$, $f_{\bar
q}=1/\left(e^{q_2^\mu u_\mu/T}+1\right)$ and $f_g=1/\left(e^{k^\mu
u_\mu/T}-1\right)$. By using Bjorken's hydrodynamics~\cite{Bjorken},
the fluid velocity $u_\mu$ appeared in the distributions and the
temperature $T$ in the distributions and gluon mass are functions of
coordinates $({\bf x}_t,\eta)$ at fixed ${\bf b}$ and determined by
the ideal hydrodynamic equations~\cite{Zhu}
\begin{eqnarray}
\label{hydro}
\partial_{\tau}E+\nabla\cdot{\bf M}&=&-(E+p)/\tau,\nonumber\\
\partial_{\tau}M_x+\nabla\cdot(M_x{\bf v})&=&-M_x/\tau-\partial_x p,\nonumber\\
\partial_{\tau}M_y+\nabla\cdot(M_y{\bf v})&=&-M_y/\tau-\partial_y p,\nonumber\\
\partial_{\tau}R+\nabla\cdot(R{\bf v})&=&-R/\tau
\end{eqnarray}
with the Lorentz factor $\gamma=1/\sqrt{1-{\bf v}^2}$ and
definitions $E=(\epsilon+p)\gamma^2-p$, ${\bf
M}=(\epsilon+p)\gamma^2{\bf v}$ and $R=\gamma n$ as functions of
energy density $\epsilon$, pressure $p$ and baryon density $n$ of
the medium. To close the hydrodynamic equations, we need a equation
of state to describe the nature of the QGP. We take the result from
the lattice QCD simulation with a phase transition temperature of
deconfinement $T_c=190$ MeV~\cite{Bazavov}. Since the colliding
energy is so high in heavy ion collisions at RHIC and LHC, the net
baryon density in the fireball is rather small~\cite{Qm2011}, we can
simply set $n=0$ in the numerical calculations. The initial
condition of the hydrodynamic equations is controlled by the nuclear
geometry which determines the impact parameter and the colliding
energy which governs the ratio of soft to hard
contributions~\cite{Heinz}.

Since we did not consider the initial dimuonium production before
the QGP formation and have neglected the loss term in the transport
equation (\ref{transport}), its analytic solution becomes simple,
\begin{eqnarray}
\label{solution}
f({\bf p}_t,y,{\bf x}_t,\eta,\tau|{\bf
b}) &=& \int^{\tau}_{\tau_0}d\tau'{\beta\left({\bf p}_t,y,{\bf
X}_t(\tau'),H(\tau'),\tau'|{\bf b}\right)\over \Delta(\tau')}\nonumber\\
& &\times\Theta(T({\bf X}_t,H,\tau'|{\bf b})-T_c)
\end{eqnarray}
with the definitions
\begin{eqnarray}
\label{define}
{\bf
X}_t(\tau')&=&{\bf x}_t-{\bf v}_t\left[\tau\cosh(y-\eta)
-\tau'\Delta(\tau')\right],\nonumber\\
H(\tau')&=&y-\textrm{arcsinh}\left(\tau/\tau'\textrm{sinh}(y-\eta)\right),\nonumber\\
\Delta(\tau')&=&\sqrt{1+(\tau/\tau')^2\textrm{sinh}^2(y-\eta)},
\end{eqnarray}
where the local temperature $T({\bf X}_t,H,\tau'|{\bf b})$ as a
function of time and coordinates at fixed impact parameter is
determined by the evolution of the medium (\ref{hydro}), the
step function $\Theta$ indicates that the mechanism of generating dimuonium
discussed here can only take place in deconfined region,
and the coordinate shifts ${\bf x}_t \to {\bf X}_t$ and $\eta \to
H$ in the solution (\ref{solution}) reflect the leakage effect in
the transverse and longitudinal directions. The time integration is from the initial
time $\tau_0$ to $\tau$. By integrating
the distribution over the phase space, we obtain the dimuonium transverse momentum
distribution at fixed impact parameter ${\bf b}$. For the finally observed
dimuonium distribution, $\tau$ should be so chosen that it is not earlier than the end time $\tau_f$ of the QGP.
\begin{eqnarray}
\label{transverse}
f(p_t|{\bf b})&=&\frac{dN({\bf b})}{2\pi
p_tdp_t}\nonumber\\
&=&\frac{\tau_c}{(2\pi)^3}\int d^2{\bf
x}_tdyd\eta E_t\cosh(y-\eta)\nonumber\\
&\times&f({\bf p}_t,y,{\bf x}_t,\eta,\tau_c|{\bf b}),
\end{eqnarray}
where $\tau_c$ is an arbitrary time after $\tau_f$, since the momentum distribution is unchanged for all $\tau > \tau_f$.

\begin{figure}[!hbt]
\centering
\includegraphics[width=0.45\textwidth]{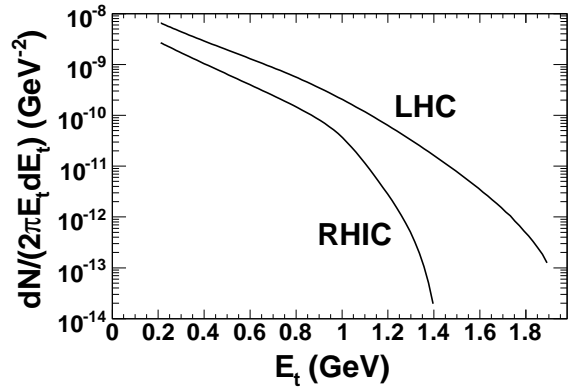}
\caption{The dimuonium transverse energy distribution $dN/(2\pi
E_tdE_t)$ in central Au+Au collisions at RHIC energy $\sqrt
{s_{NN}}=200$ GeV and Pb+Pb collisions at LHC energy $\sqrt
{s_{NN}}=5.5$ TeV. } \label{fig2}
\end{figure}

We can define the transverse energy distribution $f(E_t|{\bf
b})=dN({\bf b})/(2\pi E_tdE_t)$. It is shown in Fig.\ref{fig2} for
central (b=0) Au+Au collisions at RHIC energy $\sqrt {s_{NN}}=200$
GeV and Pb+Pb collisions at LHC energy $\sqrt {s_{NN}} = 5.5$ TeV.
We have taken the effective coupling constant $\alpha_s=0.3$
(corresponding to $T/T_c\simeq 1.5-2$~\cite{Kaczmarek}) and the
dimuonium mass $m_d\simeq 2m_\mu=211$ MeV (neglecting the binding
energy $1.4$ KeV) in the numerical calculations. While the
dimuoniums distribute in a wider region at LHC, they behave
similarly at two energies. The result in the low $E_t$ region of
$E_t < 1$ GeV can be parameterized as a thermal distribution
$f(E_t)\sim e^{-E_t/T_{eff}}$ with a slope parameter $T_{eff}=195$
MeV at RHIC and $240$ MeV at LHC. This indicates that the
thermodynamic information of the medium carried by quarks and gluons
is partly inherited by the produced dimuoniums and can be used to
signal the QGP formation in high energy nuclear collisions.
\vspace{-0.8cm}
\begin{figure}[!hbt] \centering
\includegraphics[width=0.5\textwidth]{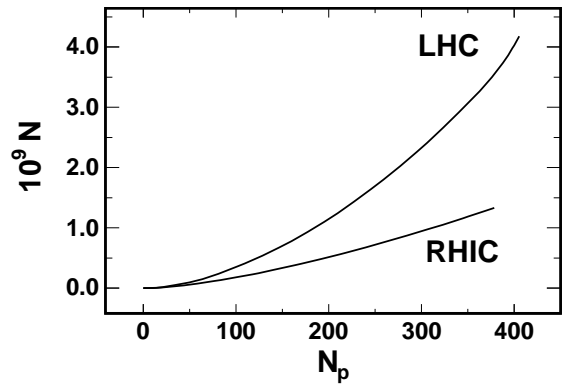}
\caption{The rescaled dimuonium number $10^9N$ as a function of
participant number $N_p$ in Au+Au collisions at RHIC energy $\sqrt
{s_{NN}}=200$ GeV and Pb+Pb collisions at LHC energy $\sqrt
{s_{NN}}=5.5$ TeV. } \label{fig3}
\end{figure}

Fig.\ref{fig3} shows the momentum integrated dimuonium yield $N({\bf
b})=2\pi\int f(p_t|{\bf b})p_tdp_t$ as a function of the number of
participant nucleons $N_p$ in heavy ion collisions at RHIC and LHC
energies. The relation between $N_p$ and the impact parameter ${\bf
b}$ can be easily determined by the nuclear geometry. With
increasing centrality, the participant number increases, and the
temperature, the life time and the space region of the formed QGP
increase. As a result, the dimuonium yield goes up with centrality,
due to the enhancement of the quark and gluon numbers. For central
collisions with maximum $N_p$, the dimuonium number is $1.3\times 10^{-9}$
at RHIC energy and becomes about
$3$ times larger at LHC energy, as shown in Fig.\ref{fig3}.

Let's now compare the two dimuonium production mechanisms in heavy
ion collisions. One is through the pure electromagnetic channel,
$A_1A_2\rightarrow A_1A_2(\mu^+\mu^-)$, proposed by Ginzburg et
al.~\cite{Ginzburg}, and the other is inside the formed QGP, $q\bar
q\rightarrow (\mu^+\mu^-)g$, discussed here. For impact parameter
$b>2R_A$, where $R_A$ is the colliding nuclear radius, there is no
QGP formed, the production is only through the electromagnetic
channel. However, for $b < 2R_A$, the colliding nuclei are broken,
and the assumption of the replacement of the perturbation parameter
$\alpha$ by $Z\alpha$ with each photon exchange is no longer valid
for the electromagnetic channel. For $b\ll 2R_A$, the QGP is formed
with a high temperature, long life time and large size, the process
$q\bar q\rightarrow (\mu^+\mu^-)g$ becomes dominant. Since the
pure electromagnetic channel is not related to the fireball, the
transverse energy distribution in this channel can not show the
thermodynamic behavior.

In summary, we investigated the dimuonium production in the QGP
formed in relativistic heavy ion collisions. The dimuonium motion in
the QGP is described by a transport equation with the gain term
characterized by the production process $q\bar q\rightarrow
(\mu^+\mu^-)g$, and the space-time evolution of the plasma is
controlled by ideal hydrodynamic equations. By solving the coupled
transport and hydrodynamic equations for high energy nuclear
collisions at RHIC and LHC energies, we found that while the
electrodynamics dominated dimuonium yield is not high enough, the
transverse energy distribution inherits the thermodynamic behavior
of the hot medium and can be considered as an electromagnetic probe
of the QGP.

\appendix {\bf Acknowledgement:} The work is supported by the NSFC
(Grant Nos. 10975084 and 11079024) and RFDP (Grant No.20100002110080
). PZ thanks Prof. Huanzhong Huang for the stimulating discussions
in the beginning of the work.

\end{document}